\documentstyle[12pt]{article}

\begin{document}
\begin{titlepage}
\begin{center}
{\Large\bf A New Evaluation of Polarized Parton Densities
\vskip 0.6cm
in the Nucleon }\\
\end{center}
\vskip 2cm
\begin{center}
{\bf Elliot Leader}\\
{\it Imperial College, University of London\\
London WC1E 7HX, England\\
E-mail: e.leader@ic.ac.uk}
\vskip 0.5cm
{\bf Aleksander V. Sidorov}\\
{\it Bogoliubov Theoretical Laboratory\\
Joint Institute for Nuclear Research\\
141980 Dubna, Russia\\
E-mail: sidorov@thsun1.jinr.ru}
\vskip 0.5cm
{\bf Dimiter B. Stamenov \\
{\it Institute for Nuclear Research and Nuclear Energy\\
Bulgarian Academy of Sciences\\
blvd. Tsarigradsko Chaussee 72, Sofia 1784, Bulgaria\\
E-mail: stamenov@inrne.bas.bg }}
\end{center}

\vskip 0.3cm
\begin{abstract}
We present a new leading and next-to-leading QCD analysis of the world
data on inclusive polarized deep inelastic scattering.
A new set of polarized parton densities is extracted
from the data and the sensitivity of the results to the newly incorporated
SLAC/E155 proton data is discussed.\\

\end{abstract}
\vskip 0.5 cm
\end{titlepage}
\newpage
\setcounter{page}{1}

\section{Introduction}

Deep inelastic scattering (DIS) of leptons on nucleons has
remained the prime source of our understanding of the internal
partonic structure of the nucleon and one of the key areas for
the testing of perturbative QCD. Decades of experiments on
unpolarized targets have led to a rather precise determination of
the unpolarized parton densities. Spurred on by the famous EMC
experiment \cite{EMC} at CERN in 1988, there has been a huge
growth of interest in {\it polarized} DIS experiments which yield
more refined information about the partonic structure of the
nucleon, {\it i.e.}, how the nucleon spin is divided up among its
constituents, quarks and gluons. Many experiments have been
carried out at SLAC, CERN and DESY.

In this paper we present an updated version of our NLO polarized
parton densities in both the $\rm \overline{MS}$ and the JET (or
so-called chirally invariant) \cite{JET} factorization schemes
as well as the LO ones determined from the world
data \cite{EMC,data,E155p} on inclusive polarized DIS. Comparing
to our previous analysis \cite{spin99}:

~~i) For the axial charges $a_3$ and $a_8$ their updated values
are used \cite{PDG,AAC}:
\begin{equation}
a_3=g_{A}=\rm {F+D}=1.2670~\pm~0.0035,~~~a_8=3\rm {F-D}=0.585~\pm~0.025~.
\label{ga,3FD}
\end{equation}

~ii) In our ansatz for the input polarized parton densities
\begin{equation}
\Delta f_i(x,Q^2_0)=\rho_i(x)~f_i^{\rm MRST}(x,Q^2_0)
\label{input}
\end{equation}
we now utilize the MRST'99 set \cite{MRST99} of unpolarized
parton densities $f_i(x,Q^2_0)$ instead of the MRST'98 one.

iii) The recent SLAC/E155 proton data \cite{E155p} are
incorporated in the analysis.

~iv) The positivity constraints on the polarized parton densities
are discussed.

\section{Method of Analysis}

The nucleon spin-dependent structure function of interest,
$g^N_1(x,Q^2)$, is a linear combination of the asymmetries
$A^N_{\parallel}$ and $A^N_{\bot}$ (or the related virtual
photon-nucleon asymmetries $A^N_{1,2}$) measured with the target
polarized longitudinally or perpendicular to the lepton beam,
respectively. Neglecting as usual the subdominant contributions,
$~A_1^{N}(x,Q^2)~$ can be expressed via the polarized structure
function $~g_1^{N}(x,Q^2)~$ as
\begin{equation}
A_1^{N}(x,Q^2)\cong (1+\gamma^2){g_1^{N}(x,Q^2)\over
F_1^{N}(x,Q^2)}
\label{assym}
\end{equation}
or
\begin{equation}
A_1^{N}(x,Q^2)\cong {g_1^{N}(x,Q^2)\over
F_2^{N}(x,Q^2)}2x[1+R^{N}(x,Q^2)]
\label{assymF2R}
\end{equation}
using the relation between the unpolarized structure function
$F_1(x,Q^2)$ and the usually extracted from unpolarized DIS
experiments $F_2(x,Q^2)$ and $R(x,Q^2)$
\begin{equation}
2xF^N_1 = F_2^N(1+\gamma^2)/(1 + R^N)~~~~~(N=p,n,d).
\label{ratio}
\end{equation}
In Eq. (\ref{assym}) the kinematic factor $\gamma^2$ is given by
\begin{equation}
\gamma^2={4M^2_{N}x^2\over Q^2}~.
\label{g2}
\end{equation}
In (\ref{g2}) $\rm M_N$ is the nucleon mass. It should be noted that in the
SLAC and HERMES kinematic region $\gamma$ cannot be neglected.

In the NLO QCD approximation the quark-parton decomposition of the proton
structure function $g_1^{p}(x,Q^2)$ has the following form (a similar formula
holds for $g_1^{n}$):
\begin{equation}
g_1^{p}(x,Q^2)={1\over 2}\sum _{q} ^{N_f}e_{q}^2 [(\Delta q
+\Delta\bar{q})\otimes (1 + {\alpha_s(Q^2)\over 2\pi}\delta C_q)
+{\alpha_s(Q^2)\over 2\pi}\Delta G\otimes {\delta C_G\over N_f}],
\label{g1partons}
\end{equation}
where $\Delta q(x,Q^2), \Delta\bar{q}(x,Q^2)$ and $\Delta
G(x,Q^2)$ are quark, anti-quark and gluon polarized densities in the proton
which evolve in $Q^2$ according to the spin-dependent NLO DGLAP
equations. $\delta C_{q,G}$ are the NLO terms
in the spin-dependent Wilson coefficient functions and
the symbol $\otimes$ denotes the usual convolution in Bjorken $x$
space. $\rm N_f$ is the number of flavours.

\def\thefootnote{\dagger}
It is well known that at NLO and beyond, the parton densities as well as
the Wilson coefficient functions become
dependent on the renormalization (or factorization) scheme
employed.\footnote{Of course, physical quantities such as the
virtual photon-nucleon asymmetry $A_1(x,Q^2)$ and the polarized
structure function $g_1(x,Q^2)$ are independent of choice of the
factorization convention.} Both the NLO polarized coefficient functions
\cite{WC} and the NLO polarized splitting functions (anomalous dimensions)
\cite{DGLAP} needed for the calculation of  $g_1(x,Q^2)$ in the
$\rm \overline{MS}$ scheme are well known at present. The corresponding
expressions for these quantities in the JET scheme can be found in
\cite{NLO_JET}.

All details of our approach to the fit of the data are given in
\cite{spin98}. Here we would like to emphasize only that
according to this approach first used in \cite{GRSV96}, the NLO
QCD predictions have been confronted to the data on the spin
asymmetry $A_1^{N}(x,Q^2)$, rather than on the $g_1^{N}(x,Q^2)$.
The choice of $A^N_1$ appears to minimize the higher twist (HT)
contributions to $g^N_1$ which are expected to partly cancel with
those of $F^N_1$ in the ratio (\ref{assym}), allowing use of data
at lower $Q^2$ (in polarized DIS most of the small $x$
experimental data points are at low $Q^2$). Indeed, we have found
\cite{LomConf} that if for $g_1$ and $F_1$ {\it leading-twist}
(LT) QCD expressions are used in (\ref{assym}), the HT
corrections $h(x)$ to $A_1(x,Q^2)=A_1(x,Q^2)_{\rm LT}+h(x)/Q^2$,
extracted from the data, are negligible and consistent with zero
within the errors (see Fig. 1). (Note that the polarized parton
densities in QCD are related to the leading-twist expression of
$g_1$.) On other hand, it was shown \cite{GRSV2000} that if $F_2$
and $R$ in Eq. (\ref{assymF2R}) are taken from experiment (as has
been done in some of the analyses) the higher twist corrections
to $A_1$ are sizeable and important. So, in order to extract the
polarized parton densities from $g_1$ data the HT contribution to
$g_1$ (unknown at present) has to be included into data fit. Note
that a QCD fit to the $g_1$ data keeping in $g_1(x,Q^2)_{QCD}$
only the leading-twist expression leads to some "effective"
parton densities which involve in themselves the HT effects and
therefore, are not quite correct. These results suggest that in
order to determine polarized parton densities less sensitive to
higher twist effects, it is preferable at present to analyze
$A_1$ data directly using for $g_1$ and $F_1$ their leading twist
expressions. Bearing in mind this discussion one must be careful
when comparing the polarized parton densities determined from the
inclusive DIS data with those extracted from semi-inclusive DIS
data \cite{SIDIS}. The results will depend, especially in a LO
QCD analysis, upon whether the higher twist terms are or are not
taken into account.\\

Following the procedure of our previous analyses we have extracted the
NLO (as well as LO) polarized parton densities from the fit to the world
data on $A_1^N(x,Q^2)$ using for the flavour non-singlet combinations of
their first moments
\begin{equation}
a_3 = (\Delta u+\Delta\bar{u})(Q^2) - (\Delta d+\Delta\bar{d})(Q^2)~,
\label{a3ga}
\end{equation}
\begin{equation}
a_8 =  (\Delta u +\Delta\bar{u})(Q^2) + (\Delta d + \Delta\bar{d})(Q^2)
- 2(\Delta s+\Delta\bar{s})(Q^2)~,
\label{3FD}
\end{equation}
the sum rule values (\ref{ga,3FD}). The sensitivity of the polarized
parton densities
to the deviation of $a_8$ from its SU(3) flavour symmetric value (0.58)
has been studied and the results are given in \cite{SU3}.
Here we will present only the polarized parton densities corresponding to
the SU(3) symmetric value of $a_8$ in (\ref{ga,3FD}).

What we can deduce from inclusive DIS in the absence of charged current
neutrino data is the sum of the polarized quark and anti-quark densities
\begin{equation}
(\Delta u +\Delta\bar{u})(x,Q^2),~~~(\Delta d +\Delta\bar{d})(x,Q^2),
~~~(\Delta s +\Delta\bar{s})(x,Q^2)
\label{qqb}
\end{equation}
and the polarized gluon density $\Delta G(x,Q^2)$. The non-strange
polarized sea-quark densities
$\Delta\bar{u}(x,Q^2)$ and $\Delta\bar{d}(x,Q^2)$, as well as
the valence quark densities
$\Delta u_v(x,Q^2)$ and $\Delta d_v(x,Q^2)$:
\begin{equation}
\Delta u_v \equiv \Delta u-\Delta\bar{u},~~~~
\Delta d_v \equiv \Delta d-\Delta\bar{d}
\label{valq}
\end{equation}
cannot be determined without additional assumptions about the flavour
decomposition of the sea. Nonetheless (because of the universality of the
parton densities) they are of interest for predicting the behaviour
of other processes, like polarized $pp$ reactions, etc. That is why,
we extract from the data not only the quark densities (\ref{qqb}) and
$\Delta G(x,Q^2)$, but also the valence parts $\Delta u_v(x,Q^2)$,
$\Delta d_v(x,Q^2)$ and anti-quark densities using the assumption on the
flavour symmetric sea
\begin{equation}
\Delta u_{sea}=\Delta\bar{u}=\Delta d_{sea}=\Delta\bar{d}=
\Delta s=\Delta\bar{s}.
\label{SU3sea}
\end{equation}

For the input LO and NLO polarized parton densities at $Q^2_0=1~GeV^2$ we
have adopted a very simple parametrization
\begin{eqnarray}
\nonumber
x\Delta u_v(x,Q^2_0)&=&\eta_u A_ux^{a_u}xu_v(x,Q^2_0),\\
\nonumber
x\Delta d_v(x,Q^2_0)&=&\eta_d A_dx^{a_d}xd_v(x,Q^2_0),\\
\nonumber
x\Delta s(x,Q^2_0)&=&\eta_s A_sx^{a_s}xs(x,Q^2_0),\\
x\Delta G(x,Q^2_0)&=&\eta_g A_gx^{a_g}xG(x,Q^2_0),
\label{classic}
\end{eqnarray}
where on RHS of (\ref{classic}) we have used the MRST98 (central gluon)
\cite{MRST98} and MRST99 (central gluon) \cite{MRST99} parametrizations
for the LO and NLO($\rm \overline{MS}$) unpolarized densities, respectively.
The normalization factors $A_f$ in (\ref{classic}) are fixed such that
$\eta_{f}$ are the first moments of the polarized densities.
The first moments of the valence quark densities $\eta_u$ and $\eta_d$ are
constrained by the sum rules (\ref{ga,3FD}).
The rest of the free parameters in (\ref{classic}),
\begin{equation}
\{a_u,~a_d,~\eta_s,~a_s,~\eta_g,~a_g\},
\label{free_par}
\end{equation}
have been determined from the best fit to the data.

The guiding arguments to choose for the input polarized parton densities
the ansatz (\ref{classic}) are simplicity (not too
many free parameters) and the expectation that polarized and
unpolarized densities have similar behaviour at large $x$.
Also, such an ansatz allows easy control of the positivity condition,
which in LO QCD implies:
\begin{equation}
\vert {\Delta f_i(x,Q^2_0)}\vert \leq f_i(x,Q^2_0),~~~~
\vert {\Delta\bar{f_i}(x,Q^2_0)}\vert \leq \bar{f}_i(x,Q^2_0).
\label{pos}
\end{equation}

The constraints (\ref{pos}) are consequence of a probabilistic
interpretation of the parton densities in the naive parton model, which is
still correct in LO QCD. Beyond LO the parton densities are not physical
quantities and the positivity constraints on the polarized parton densities
are more complicated. They follow from the positivity condition for the
polarized lepton-hadron cross-sections $\Delta \sigma_i$ with the unpolarized
ones ($\vert {\Delta \sigma_i}\vert \leq \sigma_i$) and include also the
Wilson coefficient functions.
It was shown \cite{AFR}, however, that for all practical purposes it is enough
at the present stage to consider LO positivity bounds, since NLO corrections
are only relevant at the level of accuracy of a few percent.

\section{Results}

In this section we present the numerical results of our fits to
the world data on $A_1(x,Q^2)$. The data used (185 experimental points)
cover the following kinematic region:
\begin{equation}
0.005 \leq x \leq 0.75,~~~~~~1< Q^2 \leq 58~GeV^2~.
\label{kinreg}
\end{equation}

The total (statistical and systematic) errors are taken into account.
The systematic errors are added quadratically.

We prefer to discuss the NLO results of analysis in the JET scheme.
To compare our NLO polarized parton densities with those extracted by other
groups, we present them also in the usually used $\rm \overline{MS}$ scheme
using the renormalization group transformation rules for the parton
densities.

It is useful to recall the transformation rules relating the first moments
of the singlet quark density, $\Delta \Sigma(Q^2)$, and the strange sea,
$(\Delta s+\Delta\bar{s})(Q^2)$, in the JET and $\rm \overline{MS}$ schemes:

\begin{equation}
\Delta \Sigma_{\rm JET}= \Delta \Sigma_{\rm \overline{MS}}(Q^2)
+N_f{\alpha_s(Q^2)\over 2\pi}\Delta G(Q^2),
\label{delsig_J_MS}
\end{equation}
\begin{equation}
(\Delta s+\Delta\bar{s})_{\rm JET}=
(\Delta s+\Delta\bar{s})_{\rm \overline{MS}}(Q^2) +
{\alpha_s(Q^2)\over 2\pi}\Delta G(Q^2),
\label{dels_J_MS}
\end{equation}
where $\Delta G(Q^2)$ is the first moment of the polarized gluon density
$\Delta G(x, Q^2)$ (note that $\Delta G$ is the same in the factorization
schemes under consideration).

A remarkable property of the JET (and so-called Adler-Bardeen (AB)
\cite{AB}) schemes is
that the singlet $\Delta \Sigma(Q^2)$, as well as the strange sea
polarization $(\Delta s+\Delta\bar{s})(Q^2)$, are
${\it Q^2}$ {\it independent} quantities. Then, in these schemes
it is meaningful to directly interpret $\Delta \Sigma$ as the
contribution of the quark spins to the nucleon spin and to
compare its value obtained from DIS region with the
predictions of the different (constituent, chiral, etc.) quark
models at low $Q^2$.

It is important to mention that the difference between the
values of the strange sea polarization, obtained in the
$\rm \overline{MS}$ and JET schemes could be {\it large} if $\Delta G$
in (\ref{dels_J_MS}) is positive and large. To illustrate how large it can
be, we present the values of $(\Delta s+\Delta\bar{s})$ at $Q^2=1~GeV^2$
obtained in our analysis of the world DIS data in the
$\rm \overline{MS}$ and JET schemes ($\Delta G=0.68$):
\begin{eqnarray}
\nonumber
(\Delta s+\Delta\bar{s})_{\rm \overline{MS}}&=& -0.13 \pm 0.04 \\
(\Delta s+\Delta\bar{s})_{\rm JET}&=&-0.07\pm 0.02.
\label{dels_numbers}
\end{eqnarray}
Note that if $\Delta G$ is larger than 0.68,
$(\Delta s+\Delta\bar{s})_{\rm JET}$ could {\it vanish} in
agreement with what is intuitively expected in quark models at
low-$Q^2$ region ($Q^2 \approx 0$).\\

The numerical results of our fits to the data are summarized in Table 1.
The best LO and NLO(JET) fits correspond to $\chi^2$ per degree of freedom
of $\chi^2_{\rm DF,LO}=0.921$ and to $\chi^2_{\rm DF,NLO}=0.871$.
In LO QCD $\Delta G(x,Q^2)$ does not contribute directly in $g_1$ and
the gluons cannot be determined from DIS data alone. For that reason the LO
fit to the data was performed using for the input polarized gluon density
$\Delta G(x,Q^2_0)$ that one extracted by the NLO fit to the data:
\begin{equation}
\Delta G(x,Q^2_0)_{\rm LO}= \Delta G(x,Q^2_0)_{\rm NLO(JET)}~.
\label{LOdelG_NLO}
\end{equation}
\def\thefootnote{\dagger}
We consider that such a procedure leads to non-realistic errors
of the rest of parameters and therefore, we present only their
central values in Table 1. It is important to note that in the
polarized case the LO approximation has some peculiarities
compared to the unpolarized one. As has been already mentioned
above, as a consequence of the axial anomaly, the difference
between NLO anti-quark polarizations $\Delta\bar {q}_i$ in
different factorization schemes could be quite large, in order of
magnitude of $\Delta\bar {q}_i$ themselves. In this case the
leading order will be a bad approximation, at least for the
polarized sea-quark densities extracted. Also, bearing in mind
that in polarized DIS most of the data points are at low $Q^2$,
lower than the usual cuts in the analyses of unpolarized data
($Q^2\geq 4-5~GeV^2$), the NLO corrections to all polarized
parton densities are large in this region and it is better to
take them into account. Nevertheless, the LO polarized parton
densities may be useful for many practical purposes: For
preliminary estimations of the cross sections in future polarized
experiments, etc. For that reason we present them in this paper,
at $Q^2=1~GeV^2$ (see Fig. 2 and Appendix) and for any $Q^2$ in
the region: $1\leq Q^2 \leq 6.10^5$ (the FORTRAN code is
available\footnote{http://durpdg.dur.ac.uk/HEPDATA/PDF}).
Further, we will discuss mainly the NLO QCD results.

As in our previous analysis \cite{spin99} a very good description of the world
data on $A_1^N$ and $g_1^N$ is achieved. The new NLO theoretical curves
for $A_1$ corresponding to the best fit practically coincide with the old
ones (see Fig. 3). The agreement with the SLAC/E155p data involved in this
analysis is also very good. This is illustrated in Fig. 4. The NLO QCD
theoretical predictions for $A_1$ corresponding to our previous analysis
of the data are also shown. They are in a good agreement with the SLAC/E155
data not available at the time our previous analysis was performed.
In Fig. 5 we compare the new NLO $g_1$ curves with the SMC and SLAC/E143
proton data. The extrapolations for $g_1^p$ in the yet unmeasured small $x$
region at different $Q^2$ are also shown. As seen from Fig. 5, the proton spin
dependent structure
function $g_1^p$ changes  sign at $x$ smaller than $10^{-3}$ and becomes
negative if the gluon polarization is positive. One of the challenges to
future polarized DIS experiments is to confirm or reject this behaviour,
quite different from the usual Regge type behaviour.

The extracted NLO(JET) polarized parton densities at $Q^2=1~GeV^2$
are shown in Figs. 6(a) and 6(b). (The explicit expressions are given in the
Appendix.) The new parton densities are found to be
within the error bands of the old ones. The positivity constraints have not
been imposed during the fit. Except for the strange sea
density $\Delta s(x)$, the polarized parton densities determined from the data
are compatible with the LO positivity bounds (\ref{pos}) imposed by the MRST99
unpolarized parton densities. However, if one uses the {\it more accurate}
LO positivity bounds on $\Delta s(x)$ obtained by using the unpolarized
strange sea density $s(x)_{BPZ}$ (Barone et al. \cite{Barone}),
$\Delta s(x)$ also lies in the allowed region. It is important to mention
that $s(x)_{BPZ}$ is determined with a higher accuracy compared to other
global fits.

In Fig. 7 we illustrate the difference between the singlet quark density
$\Delta \Sigma(x,Q^2)$ and the strange sea density
$(\Delta s + \Delta\bar{s})(x,Q^2)$ determined in the JET and
$\rm \overline{MS}$ schemes at $Q^2=1~GeV^2$. They differ essentially in
the small $x$ region and this difference increases with $\Delta G$ increasing.
In Figs. 8(a) and 8(b) we compare our NLO($\rm \overline{MS}$) polarized
parton densities with those obtained by AAC \cite{AAC} and
GRSV \cite{GRSV2000} using almost the same set of data. This comparison is
a good illustration of the present situation in polarized DIS. While
the quality and the kinematic range of the data are sufficient to determine
the valence polarized densities $\Delta u_v(x,Q^2)$ and $\Delta u_v(x,Q^2)$
with a good accuracy (if an SU(3) symmetry of the flavour decomposition of
the sea is assumed), the polarized strange quark density $\Delta s(x,Q^2)$
as well as the polarized gluon density $\Delta G(x,Q^2)$ are still weakly
constrained, especially $\Delta G$.

Finally, let us turn to the quark and gluon polarizations (the first moments
of the polarized parton densities). The results of the new as well as of the
old analysis are presented in Table 2 ($Q^2=1~\rm GeV^2$). The corresponding
values in the $\rm \overline{MS}$ scheme are also given. The changes of the
central values of the parton polarizations (JET scheme) are negligible and
within the errors of the quantities. Note that for the central value of the
axial charge $a_0(Q^2)$ (equal to $\Delta \Sigma(Q^2)_{\rm \overline{MS}}~$ in
$\rm \overline{MS}$ scheme) we obtain now somewhat smaller
value: $a_0(1~\rm GeV^2)=0.21 \pm 0.10$ than the old one: $a_0=0.26 \pm 0.10$.
The values of the LO parton polarizations are between those of the JET and
$\rm \overline{MS}$ schemes.

For the gluon polarization $\Delta G$ corresponding to the best NLO(JET) fit
we have found $\Delta G=0.68 \pm 0.32$ at $Q^2=1~\rm GeV^2$.
However, if one takes into account the sensitivity of $\Delta G$ to variation
of the non-singlet axial charge $a_8$ from its SU(3) symmetric value of 3F-D,
the positive values of $\Delta G$ could lie in the wider range [0, 1.5]
\cite{SU3}. A negative
$\Delta G$ is still {\it not} excluded from the present DIS inclusive data.

\section{Conclusion}
We have re-analyzed the world data on inclusive polarized deep inelastic
lepton-nucleon scattering in leading and next-to-leading order of QCD adding
to the old set of data the SLAC/E155 proton data. Compared to our previous
analysis: i) the updated values for the non-singlet axial charges $g_A$ and
$a_8$ and ii) in the ansatz for the input polarized parton densities
the MRST99 set of unpolarized parton densities instead of the MRST98 one
have been used. It was demonstrated that the polarized DIS data
are in an excellent agreement with the pQCD predictions for $A_1^N(x,Q^2)$
and $g_1^N(x,Q^2)$ and that the new theoretical curves practically coincide
with the old ones. A new set of NLO polarized parton densities in the JET
and $\rm \overline{MS}$ factorization schemes as well as LO polarized patron
densities have been extracted from the data.
We note that one must be careful when using LO polarized densities.
The LO QCD approximation could be a bad one, at least for the polarized
sea-quark densities, if it will turn out that the gluon polarization is
positive and large (the first direct measurement of $\Delta G$ \cite{HERMES}
as well as the QCD analyses support such a possibility).
We have found that the new NLO(JET) polarized parton densities do not change
essentially and lie within the error bands of the old parton densities.

What follows from our analysis is that the limited kinematic range and the
precision of the present generation of inclusive DIS experiments are enough
to determine with a good accuracy only the polarized parton densities
$(\Delta u +\Delta\bar{u})(x,Q^2)$ and $(\Delta d +\Delta\bar{d})(x,Q^2)$.
The polarized strange sea density $(\Delta s +\Delta\bar{s})(x,Q^2)$
as well as the polarized gluon density $\Delta G(x,Q^2)$ are still weakly
constrained, especially $\Delta G$. The non-strange polarized sea-quark
densities $\Delta\bar{u}$ and $\Delta\bar{d}$ cannot be determined, in
principle, from the inclusive DIS experiments alone without additional
assumptions. The further study of flavour decomposition of the sea as well
as a more accurate determination of the gluon polarization are important
next steps in our understanding of the partonic structure of the nucleon
and this will be done in the forthcoming and
future polarized lepton-hadron and hadron-hadron experiments.
\\

A FORTRAN package containing our NLO polarized parton densities
in both the JET and $\rm \overline{MS}$ factorization schemes as
well as LO parton densities can be found at
http://durpdg.dur.ac.uk/HEPDATA/PDF or obtained
by electronic mail:\\
sidorov@thsun1.jinr.ru or stamenov@inrne.bas.bg.

\vskip 4mm
{\large \bf Acknowledgments}
\vskip 4mm

One of us (D.S.) is grateful for the hospitality of the High Energy Section
of the Abdus Salam International Centre for Theoretical Physics, Trieste,
where this work has been completed.
This research was supported by the UK Royal Society and the JINR-Bulgaria
Collaborative Grants, by the RFBR (No 00-02-16696), INTAS 2000 (No 587) and
by the Bulgarian National Science Foundation under Contract Ph-1010.\\

\vskip 4mm
{\Large \bf Appendix}
\vskip 6mm
For practical purposes we present here explicitly our LO and NLO(JET) polarized
parton densities at $Q^2=1~\rm GeV^2$. The polarized valence quark densities
correspond to SU(3) flavour symmetric sea.  \\

LSS - LO:
\setcounter{equation}{0}
\renewcommand\theequation{A.\arabic{equation}}
\begin{eqnarray}
\nonumber
x\Delta u_v(x)&=&~~0.3112~x^{0.4222}~(1-x)^{3.177}~(~1-0.4085~x^{1/2}+
17.60~x~)~,\\
\nonumber
x\Delta d_v(x)&=&-0.01563~x^{0.2560}~(1-x)^{3.398}~(~1+37.25~x^{1/2}+
31.14~x~)~,\\
\nonumber
x\Delta s(x)&=&-0.08548~x^{0.5645}~(1-x)^{8.653}~(~1-0.9052~x^{1/2}+
11.53~x~)~,\\
x\Delta G(x)&=&~~19.14~x^{1.118}~(1-x)^{6.879}~(~1-3.147~x^{1/2}+
3.148~x~)~.
\label{LSSLO}
\end{eqnarray}
\vskip 1cm

LSS - NLO(JET):
\begin{eqnarray}
\nonumber
x\Delta u_v(x)&=&~~0.5052~x^{0.6700}~(1-x)^{3.428}~(~1+2.179~x^{1/2}+
14.57~x~)~,\\
\nonumber
x\Delta d_v(x)&=&-0.01852~x^{0.2704}~(1-x)^{3.864}~(~1+35.47~x^{1/2}+
28.97~x~)~,\\
\nonumber
x\Delta s(x)&=&-0.1525~x^{1.332}~(1-x)^{7.649}~(~1+3.656~x^{1/2}+
19.50~x~)~,\\
x\Delta G(x)&=&~~19.14~x^{1.118}~(1-x)^{6.879}~(~1-3.147~x^{1/2}+
3.148~x~)~.
\label{LSSNLO}
\end{eqnarray}

\newpage

\vskip 0.6 cm
\begin{center}
\begin{tabular}{cl}
&{\bf Table 1.} Results of the LO and NLO fits to the world
$A_1^N$ data ($Q^2_0=1~GeV^2$).\\
&The errors shown are total (statistical and systematic). The parameters
marked\\
& by (*) are fixed.
\end{tabular}
\vskip 0.6 cm
\begin{tabular}{|c|c|c|c|c|c|c|} \hline
    Fit &~~~~~~~~LO~~~~~~~&~~~~~~NLO(JET)~~~~~~~~\\ \hline
 $\rm DF$        &  185~-~4      &     185~-~6 \\
 $\chi^2$        &  166.7        &     155.9    \\
 $\chi^2/\rm DF$ &  0.921        &     0.871   \\  \hline
 $\eta_u$        &~~0.926$^*$    &    $0.926^*$    \\
 $a_u$           &~~0.121        &~~0.253~$\pm$~0.027  \\
 $\eta_d$        &-~0.341$^*$    &    $-0.341^*$      \\
 $a_d$           &~~0.102        &~~0.000~$\pm$~0.054  \\
 $\eta_s$        &-~0.055        &-~0.036~$\pm$~0.007  \\
 $a_s$           &~~0.754        &~~1.613~$\pm$~0.429  \\
 $\eta_g$        &~~0.681$^*$    &~~0.681~$\pm$~0.141  \\
 $a_g$           &~~0.149$^*$    &~~0.149~$\pm$~0.741   \\ \hline
\end{tabular}
\end{center}
\vskip 2.0 cm

\begin{center}
{\bf Table 2.} First moments (polarizations) of polarized parton densities
at $Q^2 = 1~GeV^2$.
\vskip 0.6 cm
\begin{tabular}{|c|c|c|c|c|c|c|} \hline
 ~~Fit~~&$\Delta u + \Delta\bar{u}$ &
 $\Delta d + \Delta\bar{d}$ & $\Delta s + \Delta\bar{s}$
&$\Delta G$&$\Delta \Sigma$& $ a_0 $\\ \hline
 old/JET & 0.86$\pm$0.03  & -0.40$\pm$0.05 &
 -0.06$\pm$0.02 & 0.57$\pm$0.31 & 0.40$\pm$0.06 & 0.26$\pm$0.10 \\ \hline
 new/JET& 0.85$\pm$0.03 & -0.41$\pm$0.05 & -0.07$\pm$0.02 &
 0.68$\pm$0.32 & 0.37$\pm$0.07 & 0.21$\pm$0.10 \\
 new/${\rm \overline{MS}}$ & 0.80 & -0.47 & -0.13 & 0.68 & 0.21 & 0.21\\
\hline
 LO& 0.82 & -0.45 & -0.11 & 0.68 & 0.25 & 0.25 \\ \hline
\end{tabular}
\end{center}
\vskip 0.6cm

\newpage
\noindent
{{\bf Figure Captions}}
\vskip 3mm
\noindent
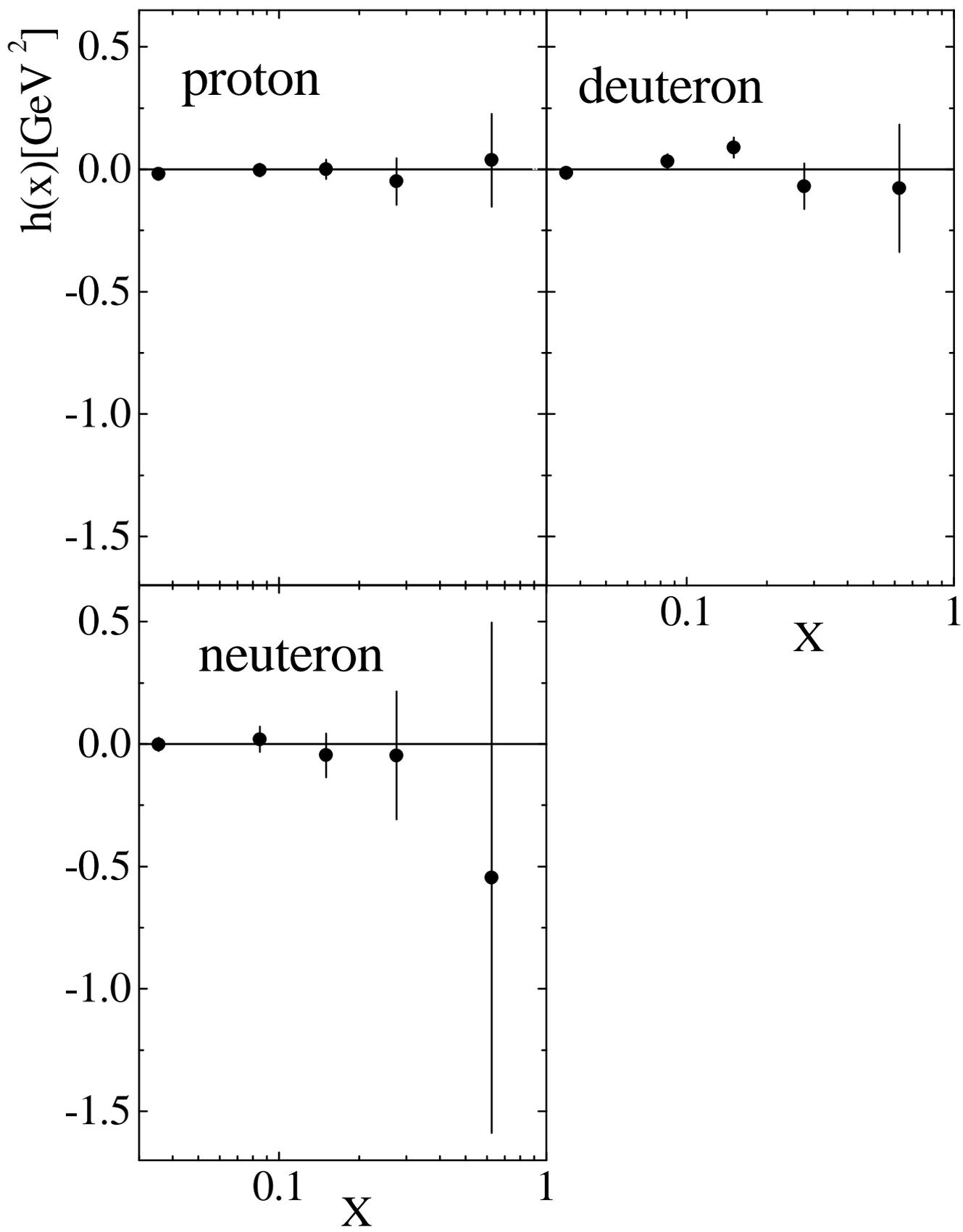
{\bf Fig. 1.} Higher twist contribution $h^N(x)$ to the
spin asymmetry $A_1^N(x,Q^2)$ extracted from the data.\\

\noindent
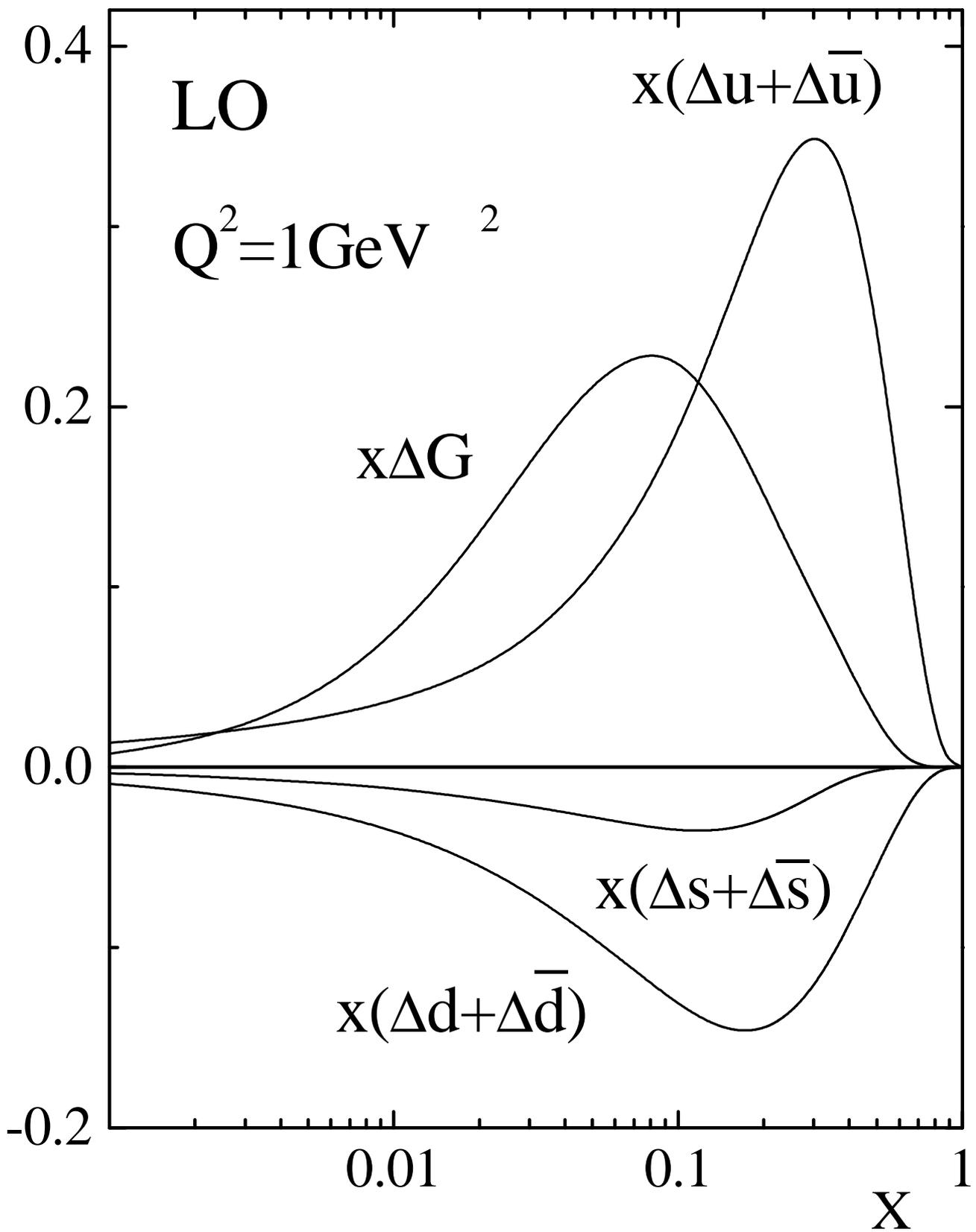
{\bf Fig. 2.} Leading order polarized parton densities at
$Q^2= 1~\rm GeV^2$.\\

\noindent
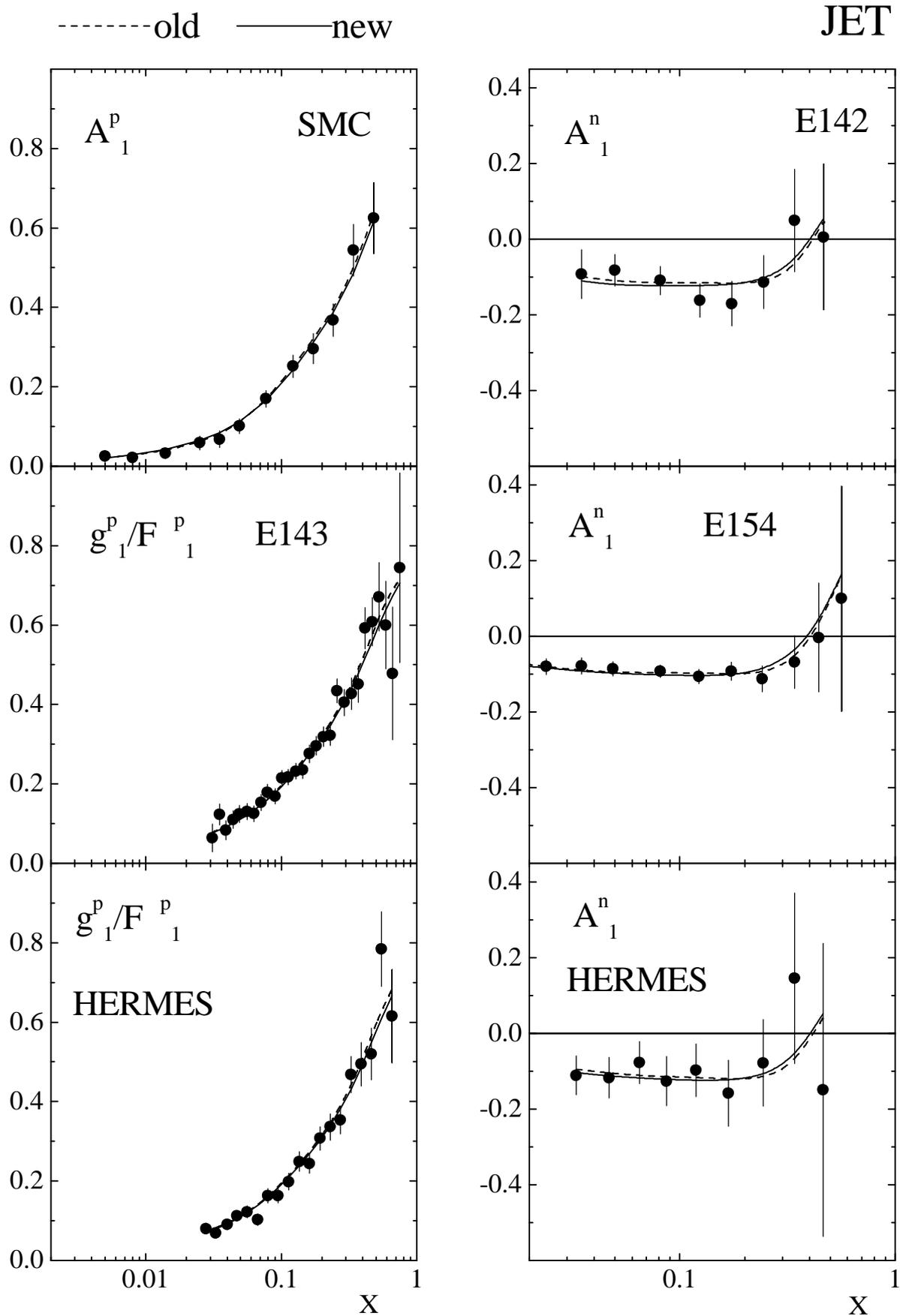
{\bf Fig. 3.} Comparison of our NLO results in JET scheme for
$~A_1^N(x,Q^2)~$ with the experimental data at the measured
$x$ and $Q^2$ values. Errors bars represent the total
(statistical and systematic) error. Our old NLO results \cite{spin99}
are shown for comparison.\\

\noindent
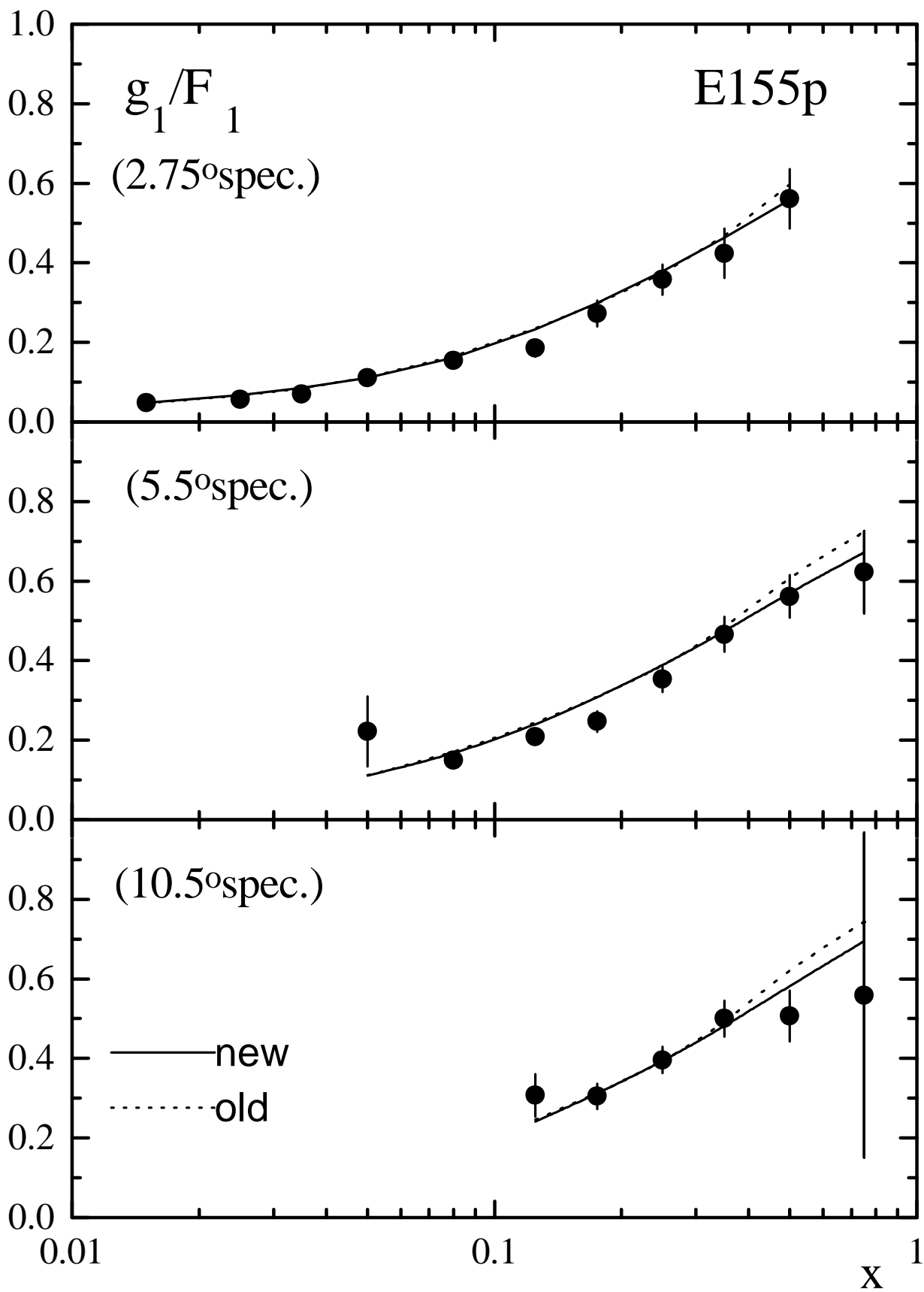
{\bf Fig. 4.} Comparison of our NLO(JET) result for $A_1^p$ with SLAC/E155p
experimental data. Error bars represent the total errors. The predictions
for $A_1^p$ (dot curves) from our old analysis \cite{spin99} are also shown.\\

\noindent
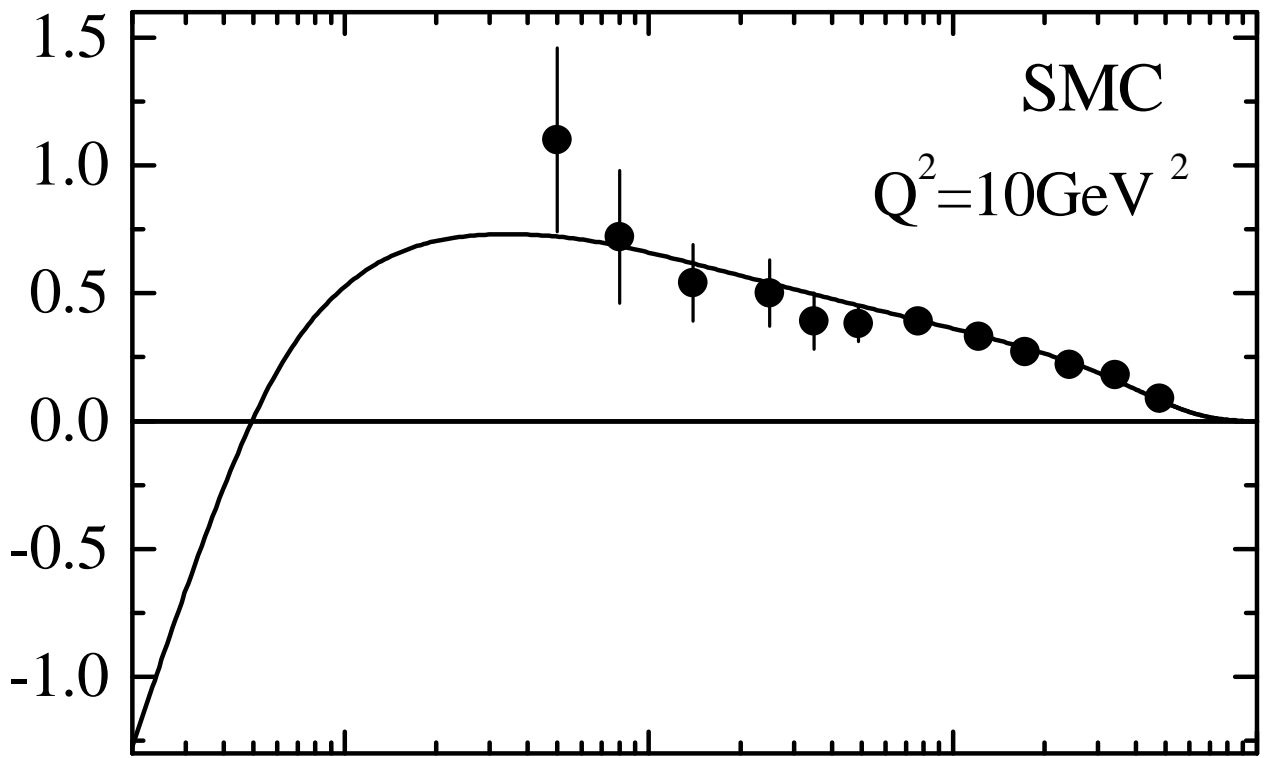
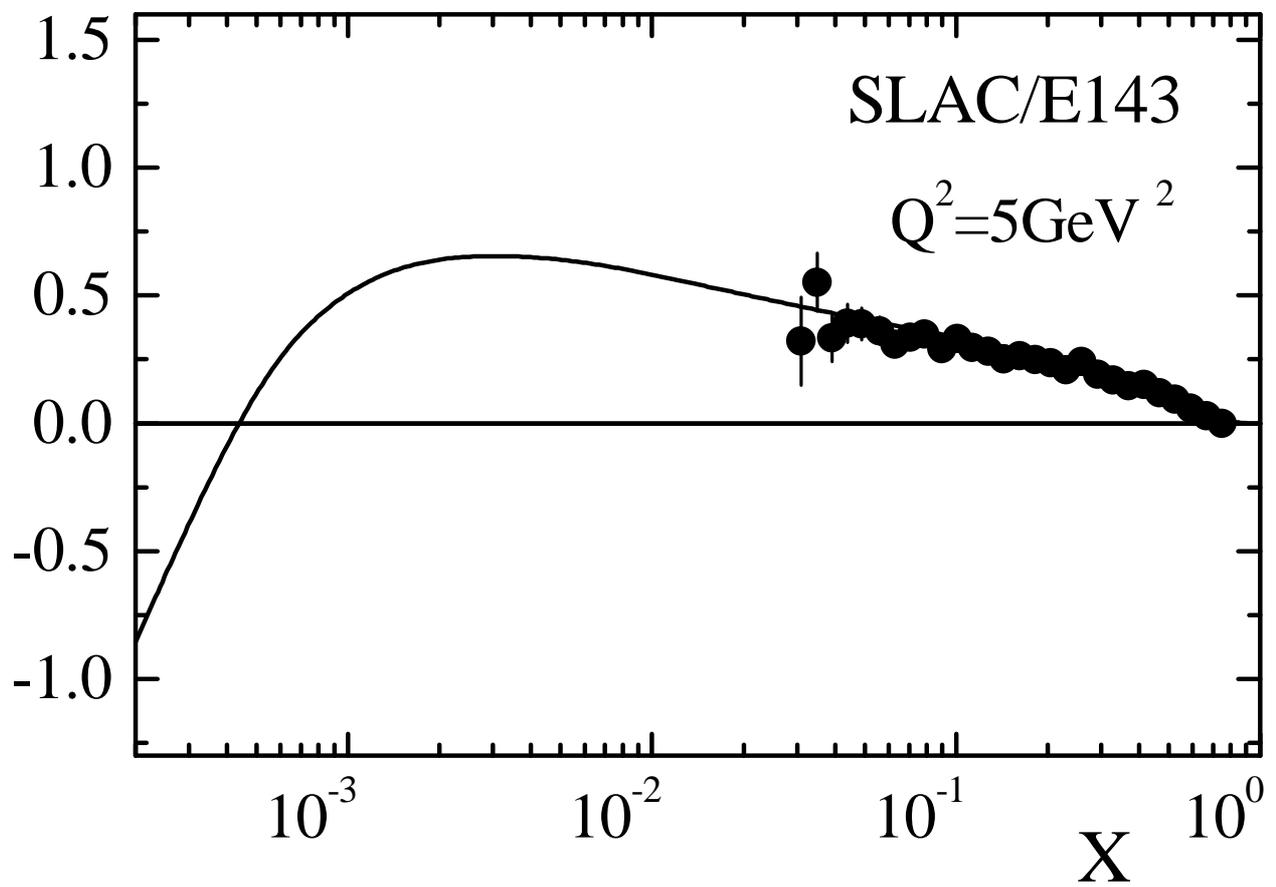
{\bf Fig. 5.} Comparison of our NLO(JET) results for $g_1^p$ with SMC and
SLAC/E143 proton data. The extrapolations at small $x$ are also shown.\\

\noindent
{\bf Fig. 6(a).} NLO(JET) polarized parton densities
$x(\Delta u+\Delta\bar{u}$) and $x(\Delta d+\Delta\bar{d}$) at
$Q^2= 1~\rm GeV^2$. The old parton densities together with their error bands
are presented for comparison.\\

\noindent
{\bf Fig. 6(b).} NLO(JET) polarized strange sea $x(\Delta s+\Delta\bar{s}$)
and gluon polarized parton densities at $Q^2= 1~\rm GeV^2$. The old parton
densities together with their error bands are presented for comparison.\\

\noindent
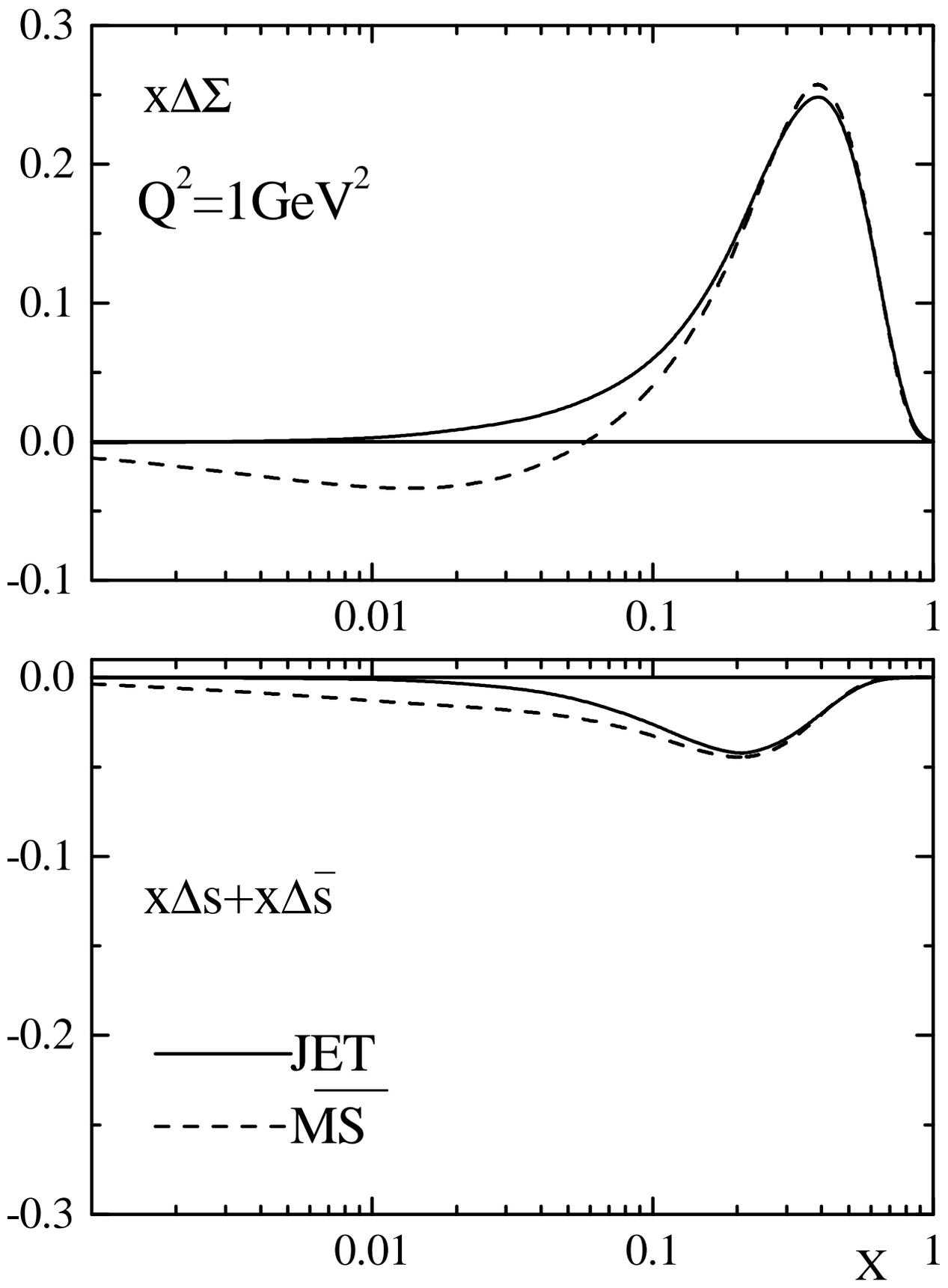
{\bf Fig. 7.} Comparison between NLO polarized singlet and strange sea parton
densities at $Q^2= 1~\rm GeV^2$ in the JET and ${\rm \overline{MS}}$
schemes.\\

\noindent
{\bf Fig. 8(a).} Comparison between our NLO(${\rm \overline{MS}}$) polarized
valence quark densities at $Q^2= 1~\rm GeV^2$ with those obtained by
AAC (NLO-2 set) \cite{AAC} and GRSV (`standard' scenario) \cite{GRSV2000}.\\

\noindent
{\bf Fig. 8(b).} Comparison between our NLO(${\rm \overline{MS}}$) polarized
strange quark and gluon densities at $Q^2= 1~\rm GeV^2$ with those obtained by
AAC (NLO-2 set) \cite{AAC} and GRSV (`standard' scenario) \cite{GRSV2000}.

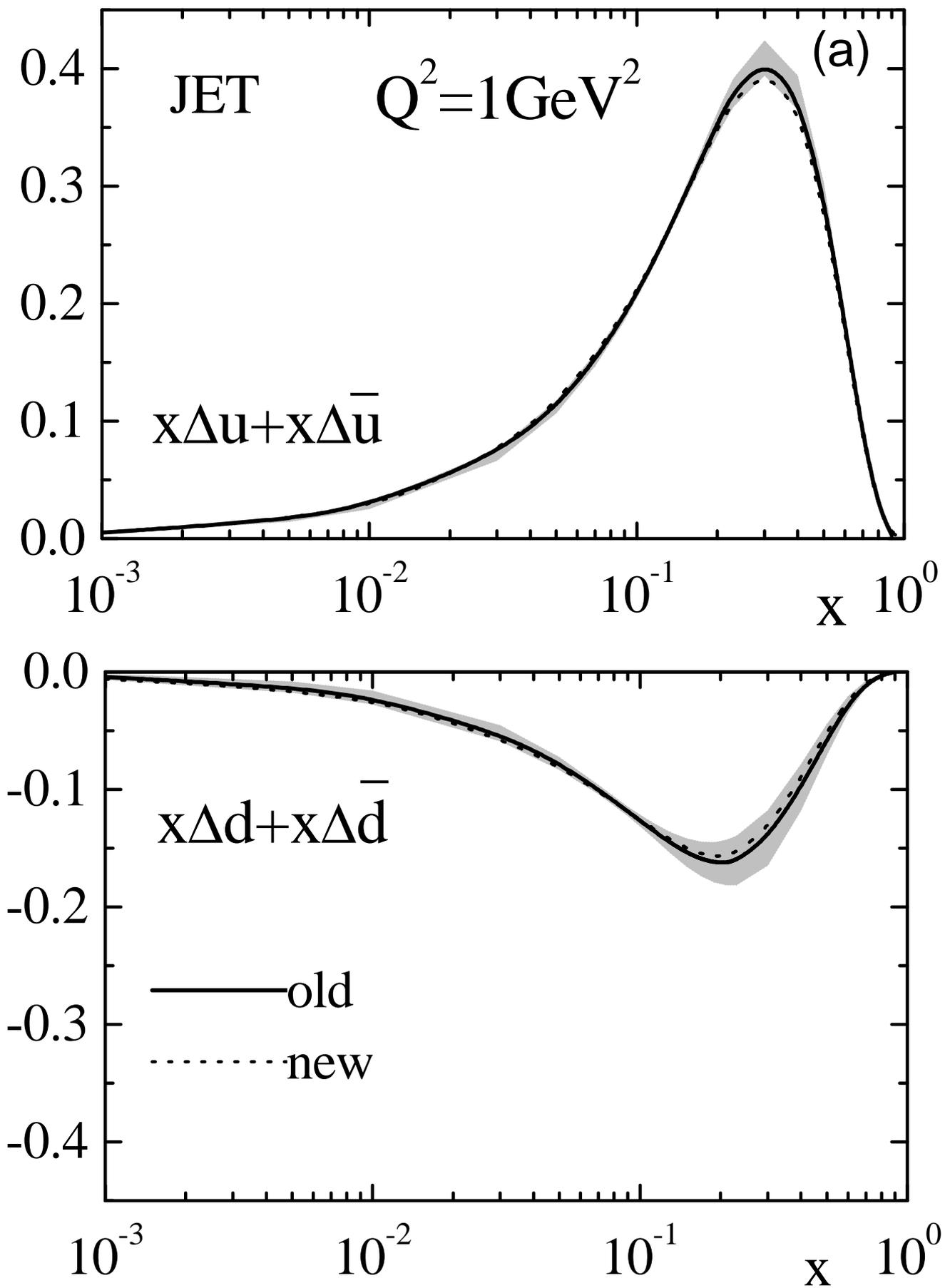
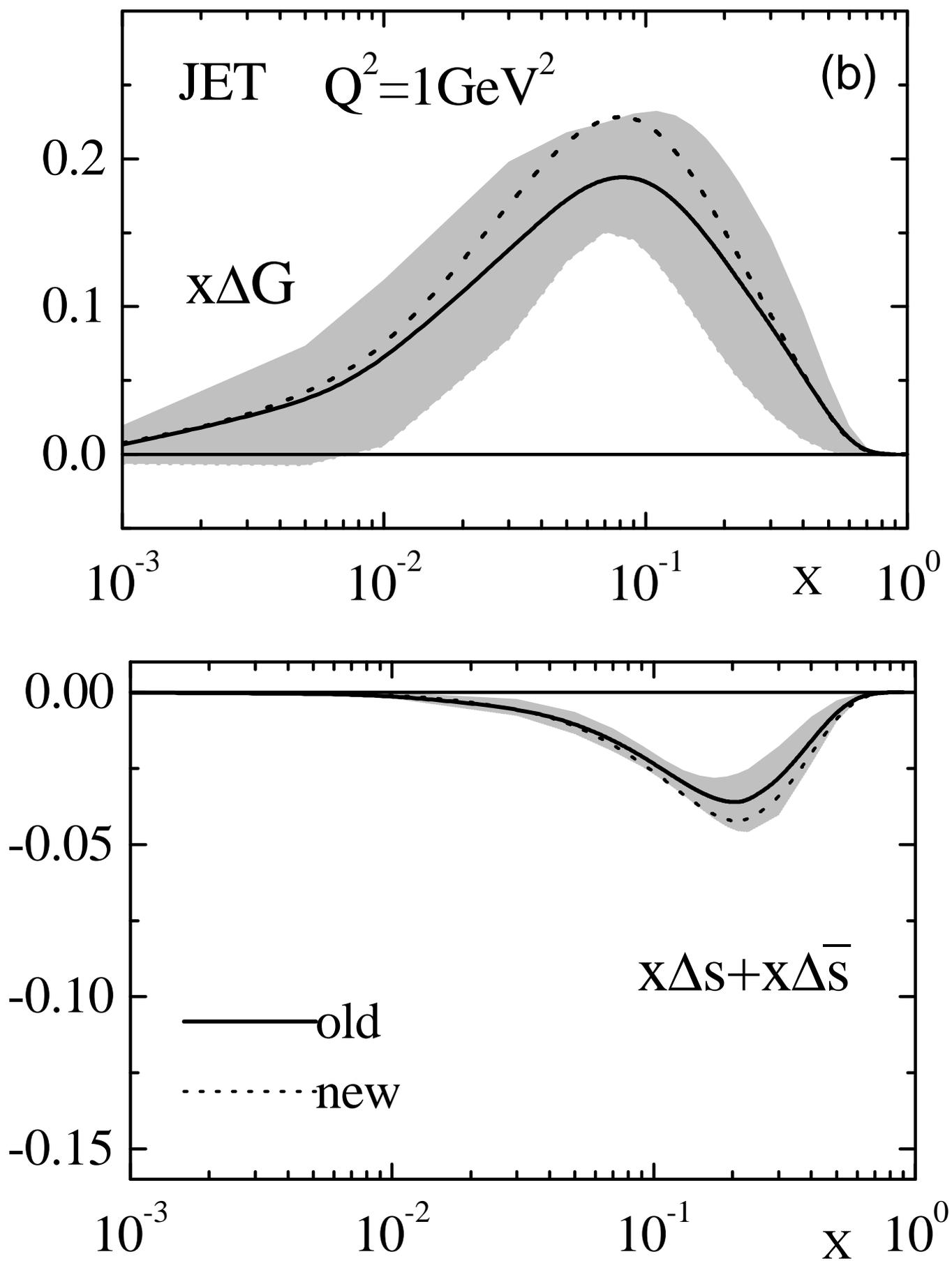
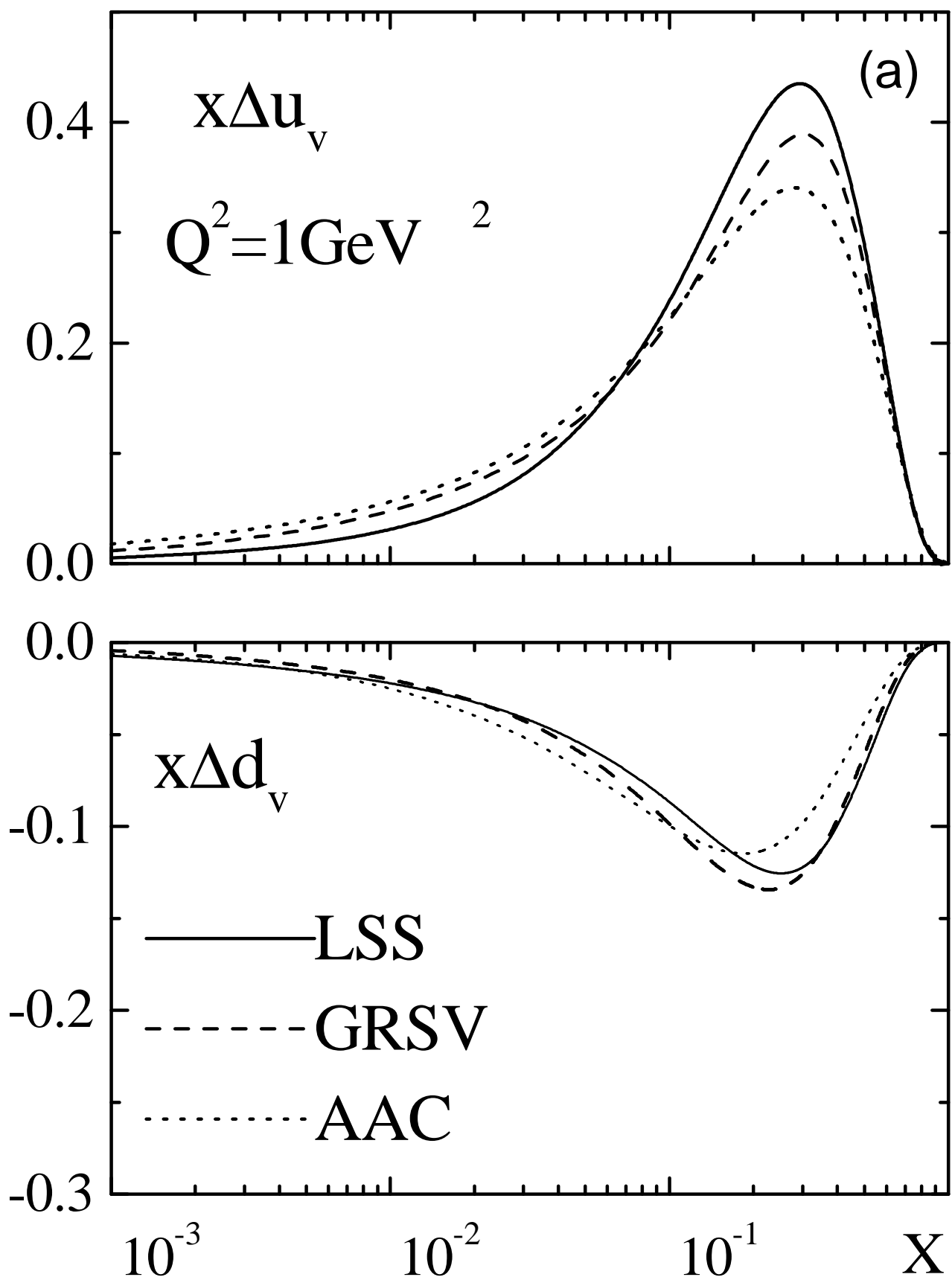
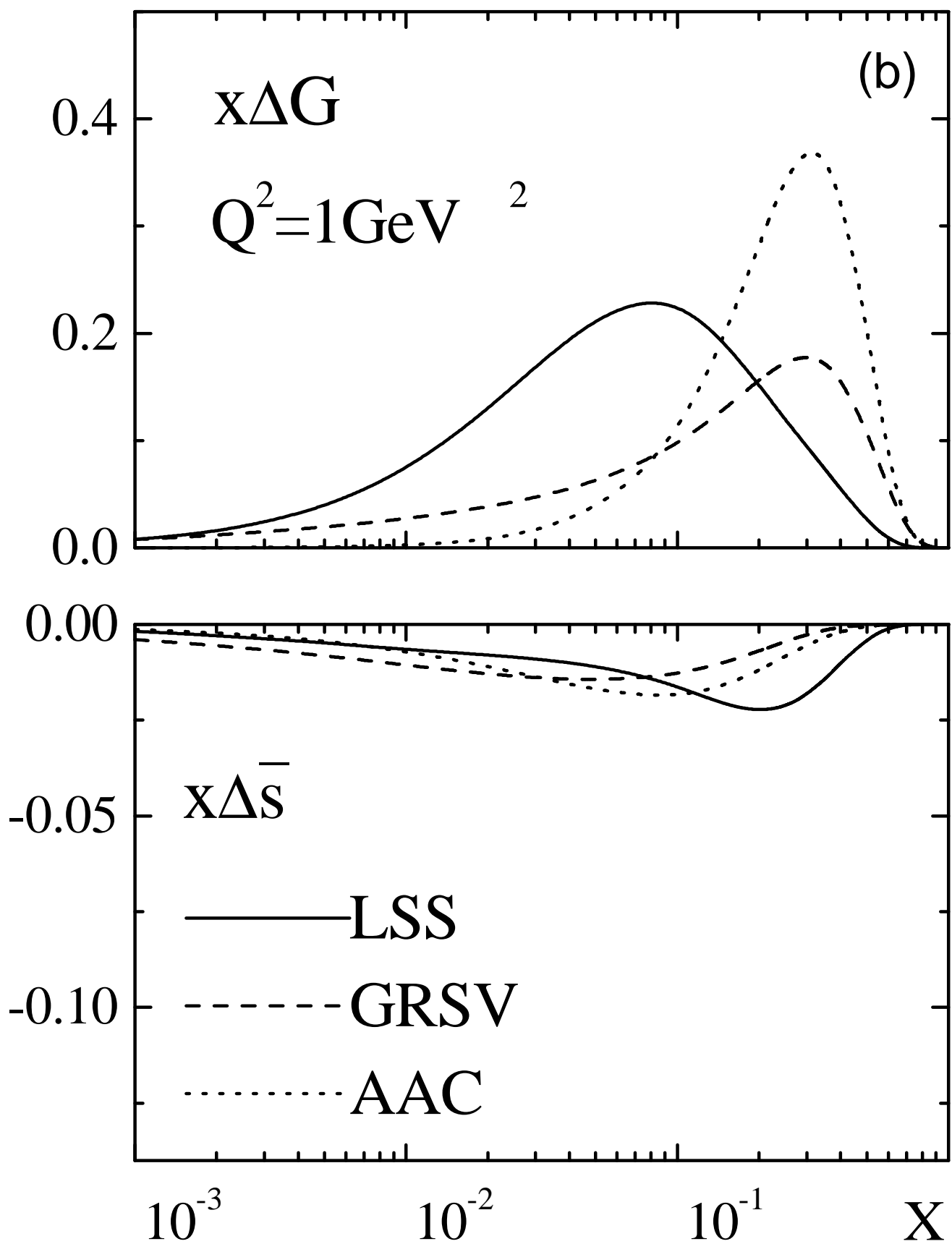
\end{document}